# Strong Ordering by Non-uniformity of Thresholds in a Coupled Map Lattice

Frode Torvund and Jan Frøyland

*Department of Physics, University of Oslo, P.O. Box 1048 Blindern, N-0316 Oslo, Norway*

## Abstract

The coupled map lattice by Olami *et al.* [Phys. Rev. Lett. **68**, 1244 (1992)] is "doped" by letting just *one* site have a threshold, $T^*_{\max}$, bigger than the others. On an $L \times L$ lattice with periodic boundary conditions this leads to a transition from avalanche sizes of about one to exactly $L^2$, and after each avalanche stresses distributes among only five distinct values, $\tau_k$, related to the parameters $\alpha$ and $T^*_{\max}$ by $\tau_k = k\alpha T^*_{\max}$ where $k = 0, 1, 2, 3, 4$. This result is independent of lattice size. The transient times are inversely proportional to the amount of doping and increase linearly with $L$.

During the last few years there has been a great amount of interest in a coupled map lattice introduced by Olami *et al.* [1]. The map can be derived from a spring-block model of an earthquake fault proposed by Burridge and Knopoff [2]. The model consists of blocks coupled to its nearest neighbours by elastic springs. Each block is exposed to a uniformly increasing stress which discharges when it reaches a certain threshold, and part of the stress is then transferred to the nearest neighbours through the couplings. The corresponding coupled map lattice is defined on an $L \times L$ lattice. Each site $(i, j)$ is associated with a stress $T_{ij}$. Initially, the stresses are randomly distributed over the lattice. The stresses are increased at a slow rate, the same rate for all sites. Finally, one site will reach the threshold, $T^*$, and topple. The following relaxation rules are then applied to the system:

$$\begin{aligned} T_{nn} &\to T_{nn} + \alpha T_{ij} \\ T_{ij} &\to 0, \end{aligned} \tag{1}$$

where $nn$ denotes nearest neighbours to the toppling site $(i, j)$. The parameter $\alpha$ controls the level of conservation in the system. One must have $0 < \alpha \leq 0.25$. The redistribution of stress may cause some of the $nn$ sites to exceed the threshold and thus start an avalanche of toppling sites.

For open boundary conditions the model displays a power law in the distribution of avalanche sizes for $0.05 \leq \alpha \leq 0.25$ [3]. The system seems to organize itself into a critical state without having to fine-tune the parameter $\alpha$, and it is therefore claimed to exhibit *self-organized criticality* [4]. This critical behaviour is dependent on boundary conditions and on the degeneracy of the system [5,6]. With periodic boundary conditions, the degenerate system goes into a state with average avalanche size only slightly bigger than one for almost all initial conditions, and it seems clear that the spatial inhomogeneity caused by open boundary conditions is responsible for the scaling in this model. The effect of the boundary has recently been studied in detail [7]. It has been conjectured that similar avalanche scaling should be observed also with other sources of spatial inhomogeneity [5].

In order to remove the degeneracy (and thereby introduce a spatial inhomogeneity) we follow Jánosi and Kertész [8] and let the thresholds be site dependent. Site dependent thresholds have also been used by Rundle and Klein [10] in a model somewhat similar to the model by Olami *et al.* However, instead of a uniform distribution we use a gaussian distribution with mean value 1.0 and standard deviation $\sigma$. This extension of the model completely changes its behaviour both for periodic and open boundaries. For periodic boundaries and for several values of $\sigma \sim 10^{-2}$, our numerical simulations show that the system enters into a state with period one and avalanche size exactly equal to $L^2$, that is, exactly all sites topple once in each avalanche. This is a behavior similar to that in models of globally coupled biological oscillators [9]. Furthermore, for $\sigma$ small, the distribution of stresses right after an avalanche exhibits only five distinct values, $\tau_k = k\alpha T^*_{\max}$ where $k = 0, 1, 2, 3, 4$ and $T^*_{\max}$ is the highest threshold in the system (Fig. 1). Only *one site* holds the largest value, $\tau_5$, and this is precisely the site which has the highest threshold. Consequently, this is the site which triggers the next avalanche. Thus, all transport of stress in the system must be in the form of "packets" of size $\alpha T^*_{\max}$, and all sites have the same stress $T^*_{\max}$ when they topple. We have not observed any case when this did not happen for lattice sizes in the range $15 \leq L \leq 50$, $0.05 \leq \alpha \leq 0.249$ and for many different initial conditions. The tendency towards clustering around four values has already been observed by Grassberger, in the limit $\alpha \to 0.25$ of the degenerate model [5] and by Zhang in a slightly different model [11]. Recently, a similar "quantization" of stress was observed by Corral *et al.* in a nonlinearly driven model with open boundary conditions [12].

The site with the largest threshold is able to trigger all the avalanches only if the difference between $T^*_{\max}$ and the stress at this site is less than the difference between threshold stress and stress at all other sites. The next highest stress value occuring in the ordered state is $\tau_3 = 3\alpha T^*_{\max}$. Thus, the relation

$$T^*_{\max} - 4\alpha T^*_{\max} < T^*_{\min} - 3\alpha T^*_{\max} \tag{2}$$

or

$$T^*_{\max} < \frac{T^*_{\min}}{1-\alpha}, \tag{3}$$

where $T^*_{\rm min}$ is the smallest threshold in the lattice, seems to be a sufficient condition for the strongly ordered period one state, and this is consistent with our numerical results. The necessary condition will depend on initial conditions. When $\sigma$ is increased such that this condition is violated, the system enters a region of more complicated states, including periodic behavior with periods larger than one and states with either a very large or an infinite period. In the latter case, the avalanche distribution function is exponentially decreasing. There seems to be no upper limit on $\alpha$, other than the conservative limit, $\alpha < 0.25$, and no lower limit on $\sigma$ for obtaining period one.

One may ask, how many period one attractors of the type described above exist for a given lattice? We cannot answer that question, but it is certainly a large number. We simulated a lattice with $L = 4$ and $\sigma = 0.01$ and 200 different initial conditions ended in 200 different final configurations — all of the type described above.

With open boundary conditions, Jánosi and Kertész found an exponential decay of the distribution function for a large, uniform spread on the thresholds [8]. For small $\sigma$, we still find a strong tendency towards very large avalanches. In this case, the distribution of stresses immediately after a big avalanche is very similar to the case with periodic boundary conditions, but instead of just five values there are now extremely narrow distributions around the four lowest $\tau_k$ values, and in addition there are a few values scattered in the intervals between the $\tau_k$ values (Fig. 2). On inspection, it turns out that the sites with these values are all on or near the boundary.

In order to demonstrate the effect more clearly, we use periodic boundary conditions and let just *one* site have a threshold, $T^*_{\rm max}$, larger than the threshold $T^*$ of the remaining sites. The only difference from the behaviour described above was somewhat longer transient times. The time dependence of the avalanche size is illustrated in Fig. 3. The period one state implies that the doped site has the stress value $\tau_4 = 4\alpha T^*_{\rm max}$ right after an avalanche, and that the stresses on the other sites are distributed among the four remaining $\tau_k$-values. It also implies that the doped site triggers all the avalanches. We let $T^* = 1$, and must then have

$$T^*_{\max} - 4\alpha T^*_{\max} < 1 - 3\alpha T^*_{\max} \tag{4}$$

or

$$1 < T^*_{\max} < \frac{1}{1-\alpha} \tag{5}$$

in order to obtain the period one state. For instance, for $\alpha = 0.1825$ we must have $1 < T^*_{\max} < 1.223...$ . For this value of $\alpha$ we obtained a period one state for $T^*_{\max} \leq 1.21$, but not for any $T^*_{\max} > 1.22$ (Fig. 4). In the latter case we observe, as for large $\sigma$, a variety of different states for different parameter values and initial conditions. In some cases, the avalanche distribution function can be fitted to a finite-size scaling hypothesis, but we have not been able to find this behavior in any range of parameter values.

When inequality (5) is fulfilled, it is implied that the distribution of thresholds in the interval $\langle(1-\alpha)T^*_{\max}, T^*_{\max}\rangle$ is irrelevant as far as the character of the distribution of thresholds are concerned. However, it may have some influence on the number of sites having each of the four possible values.

We have mainly used lattice sizes up to $50\times50$, which was also the largest used by Olami *et al.* [1] in their original paper. It has been pointed out [5,13] that the conclusions of Olami *et al.* concerning the finite-size scaling of the avalanche size distribution cannot remain correct for sufficiently large lattices. However, in the present case it is observed that for periodic boundary conditions, the essential effect of a strongly ordered final state is completely independent of lattice size (except, of course, that transient times increase with lattice size). Thus, there is no reason to suspect that larger lattices will behave differently. However, we have made a single run with a lattice size $200 \times 200$ and indeed found the same strong ordering effect.

The transient times increases as $T^*_{\max} \to 1$ from above. For $\alpha = 0.05, 0.10, 0.15$ and $0.20$, and for $T^*_{\max}$ just above 1, we find that the transient times obeys a power law

$$t_{\mathrm{tr}} \sim (T^*_{\max} - 1)^{-\mu}, \tag{6}$$

where $\mu = 1.00 \pm 0.01$ (Fig. 5). For a lattice size $15\times15$, $\mu$ is independent of $\alpha$ within errors,

and for $\alpha = 0.20$ it is also independent of $L$ for $L = 15, 25$ and 35 (Fig. 6). The transient times seems to increase linearly with $L$ (Fig. 7); the dependence on $\alpha$ is not that simple (Fig. 8).

A strongly ordered state at one value of $T^*_{\max}$ can be used to find a strongly ordered state at another value of $T^*_{\max}$ by simply shifting the stresses $k\alpha T^*_{\max}$ in accordance with the change in $T^*_{\max}$. According to Eq.( 6) the transition time to a strongly ordered state for $T^*_{\max} = 1$ (i.e. the original model) would be infinite, but using initial conditions constructed in the way described above, we find that strongly ordered states exist also in the original model.

When $T^*_{\max} \to \frac{1}{1-\alpha}$ from below, the transient times increases in a much more erratic way than what was found for the transition $T^*_{\max} \to 1$, and no simple behaviour was observed. (See Fig. 5 and 6).

Much of the behavior described above can also be found in small systems, on which we can do some analytical calculations. Consider a system of only two sites, $a$ and $b$. If $T_b$ is the state of $b$ right after $a$ has toppled, we can define the return map $R_b(T_b)$ as the state of $b$ after the next toppling of $a$. For uniform thresholds it is easy to show that $R_b(T_b) = T_b$, where the "avalanche size" $s = 1$, i.e. there exist infinitely many marginally stable period two fixed points. However, if we introduce non-uniform thresholds by letting $T^*_a = 1 + \varepsilon$ the return map will be

$$R_b(T_b) = T_b + \varepsilon(\alpha + 1), \tag{7}$$

that is, the stress of $b$ will increase with the amount $\varepsilon(\alpha + 1)$ for each subsequent toppling of $a$ and $b$. Of course, this can only continue until $T_b$ becomes larger than $1 - \varepsilon(\alpha + 1)$. Then, as $b$ topples, $T_b \to 0$ and $T_a \to \varphi$ where $\alpha < \varphi < \varepsilon(\alpha + 1) + \alpha$. For $a$ to topple next, the condition $1 + \varepsilon - \varphi < 1$ or $\varepsilon < \varphi$ must be satisfied. A sufficient condition for this is that $\varepsilon < \alpha$, the necessary condition will depend on initial conditions. If the condition is satisfied, the following toppling of $a$ will trigger $b$ and we get an "avalanche" of size $s = 2$. We define the return map $R_a(\varphi)$ as the state of $a$ after this avalanche and obtain

$$R_a(\varphi) = \alpha(1+\varepsilon)(1+\alpha) - \alpha\varphi. \tag{8}$$

Since $\varphi > \alpha$, $R_a(\varphi) < \varepsilon(\alpha+1)+\alpha$ and the next avalanche will also be of size $s = 2$. Thus, we have the general return map

$$R_a^{n+1} = \alpha(1+\varepsilon)(1+\alpha) - \alpha R_a^n. \tag{9}$$

This is a simple linear one-dimensional map with one fixed point,

$$\tilde{R}_a = \alpha(1+\varepsilon), \tag{10}$$

which is stable since $\alpha < 1$. It is the same type of period one fixed point we found for the square lattices, but now there are only two $k$-values, $k = 0, 1$. From Eq.(7) we see that

$$N(\varepsilon,\alpha) \propto \frac{1}{\varepsilon(\alpha+1)}, \tag{11}$$

where $N$ is the number of avalanches before the fixed point is reached. The global increase of stress is proportional to the time, and from Eq.(7),

$$\Delta t_{n-1} + \Delta t_n \propto 1 + \varepsilon - \alpha, \tag{12}$$

where $\Delta t_n$ is the time between the $n$-th and the $(n+1)$-th avalanche. This yields a transient time

$$t_{\mathrm{tr}} = \sum_{n=1}^{N} \Delta t_n \propto \frac{1+\varepsilon-\alpha}{\varepsilon(\alpha+1)}. \tag{13}$$

When $\varepsilon \to 0$ this gives $t_{\mathrm{tr}} \sim (T^*_{\mathrm{max}} - 1)^{-1}$, which is in good agreement with the results of the square lattices. Still, we do not find it trivial that a lattice with local coupling has qualitatively the same asymptotic behavior as a system with only two sites.

In summary, it is demonstrated that the coupled map lattice of Olami *et al.* is not robust as a model of earthquakes, since changing the threshold at just *one* site completely changes its behaviour. In a certain range of parameters, the resultant asymptotic states are highly organized, even with open boundary conditions. One may speculate if a similar doping of other degenerate type of models could have a similar effect.

FT is grateful to Kim Christensen for several useful discussions.

## FIGURES

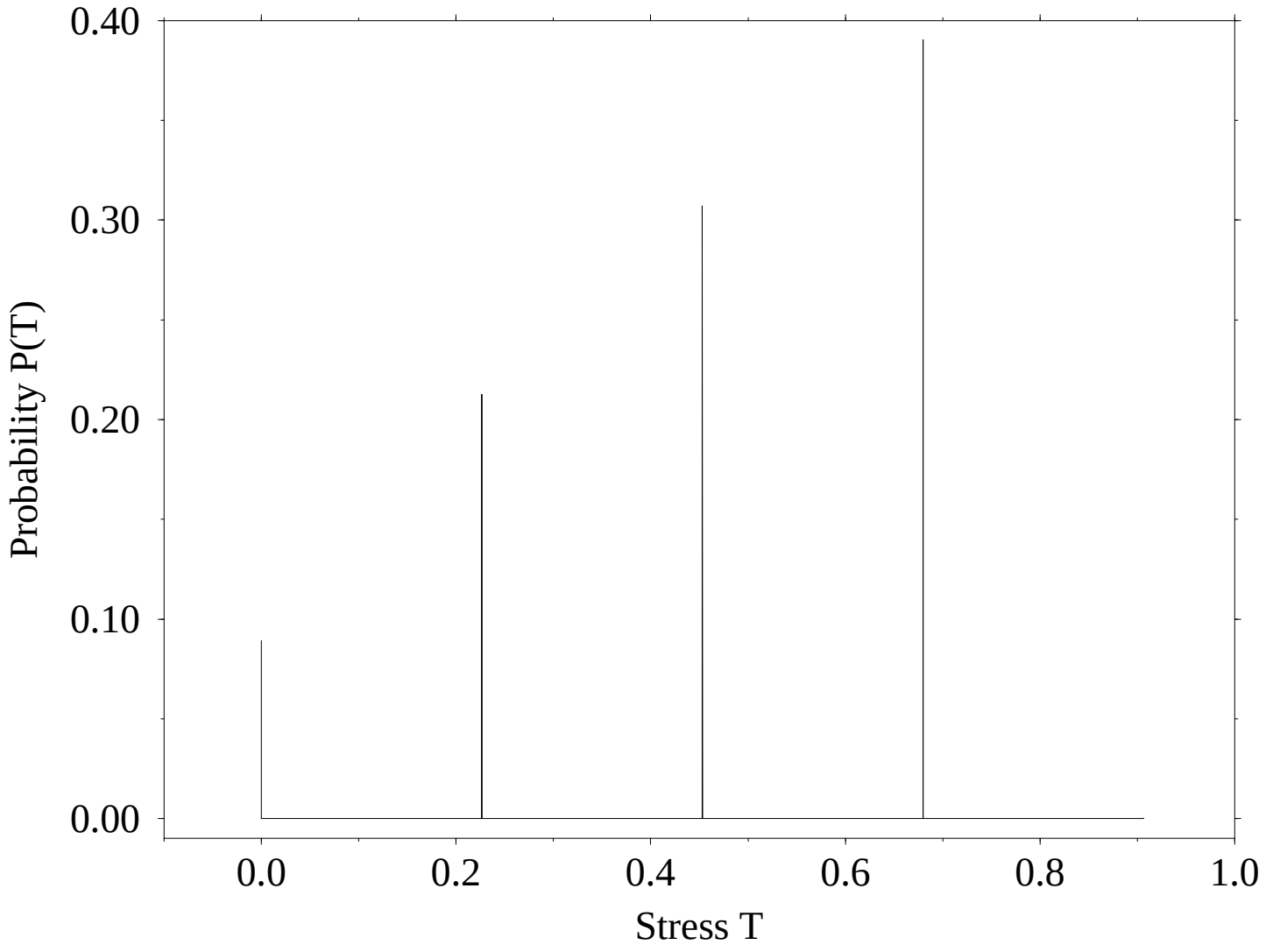

FIG. 1. Distribution of stresses immediately after an avalanche in the period one i.e. asymptotic state for a $50 \times 50$ lattice with periodic boundary conditions, $\alpha = 0.20$ and $\sigma = 0.04$.

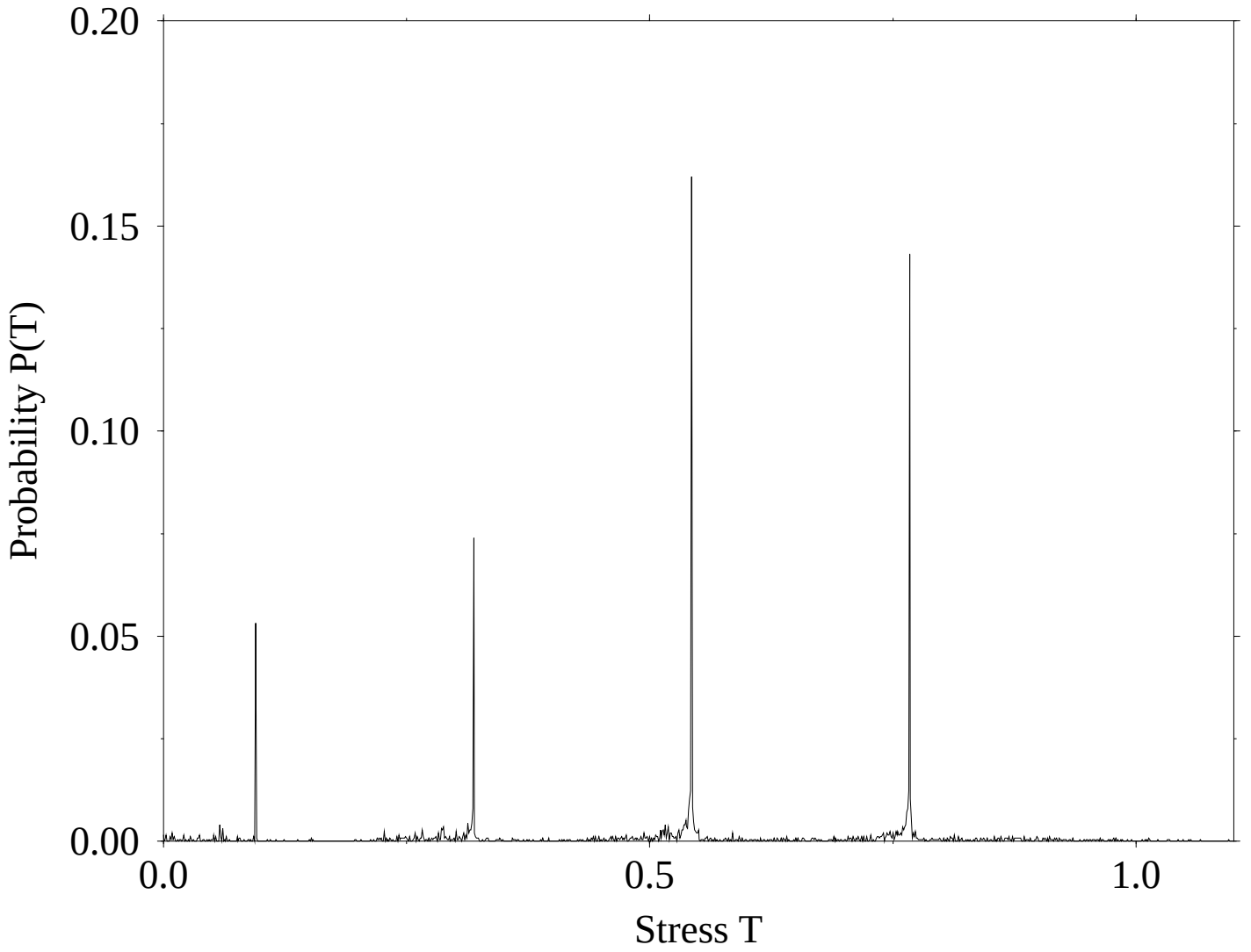

FIG. 2. Distribution of stresses after an avalanche in a quasi asymptotic state (after $1 \times 10^7$ avalanches) for a $50 \times 50$ lattice with open boundary conditions, $\alpha = 0.20$ and $\sigma = 0.04$.

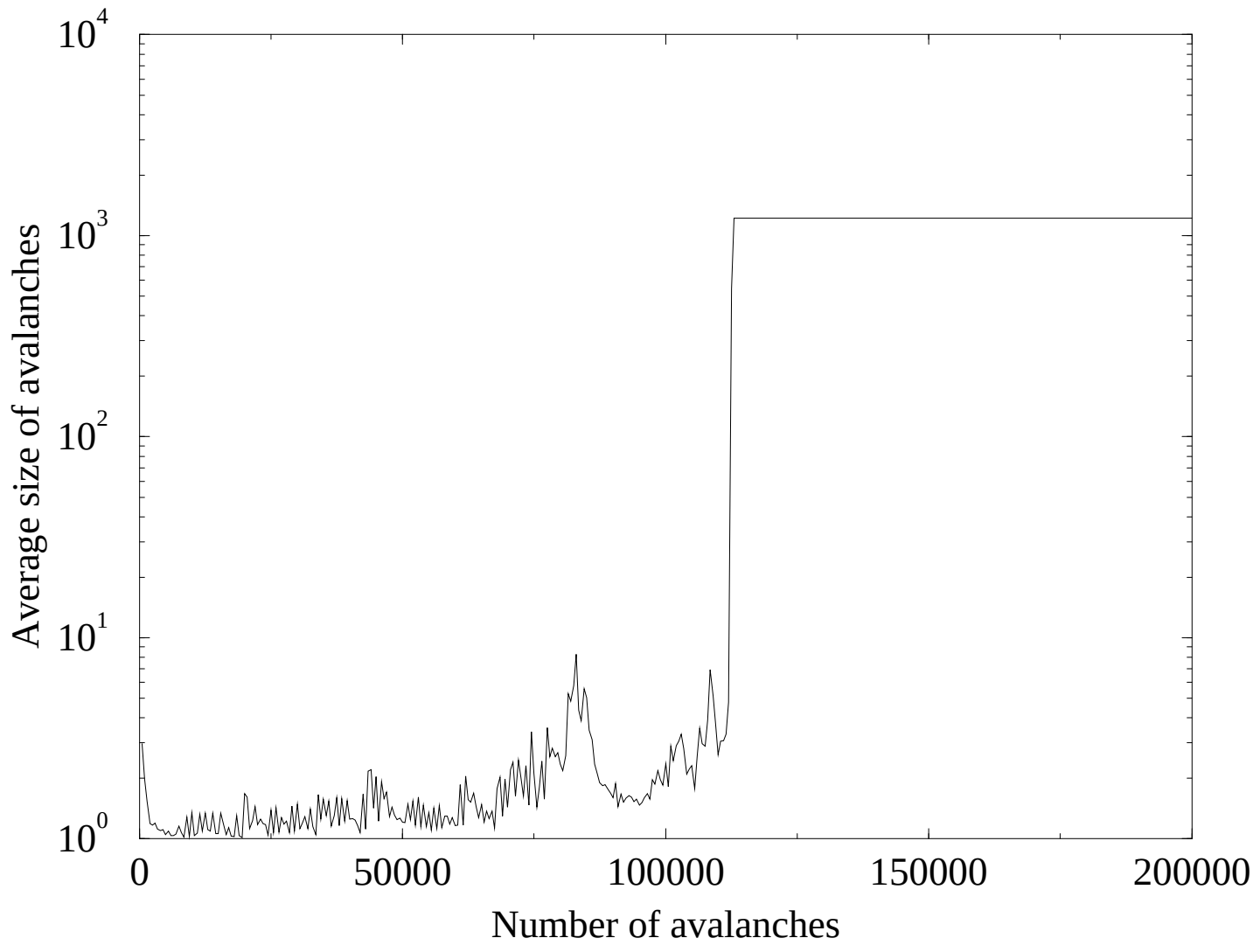

FIG. 3. Average size of the last 500 avalanches as a function of avalanche number for a $35 \times 35$ lattice with periodic boundaries and one doped site, $\alpha = 0.1825$ and $T^*_{\max} = 1.21$.

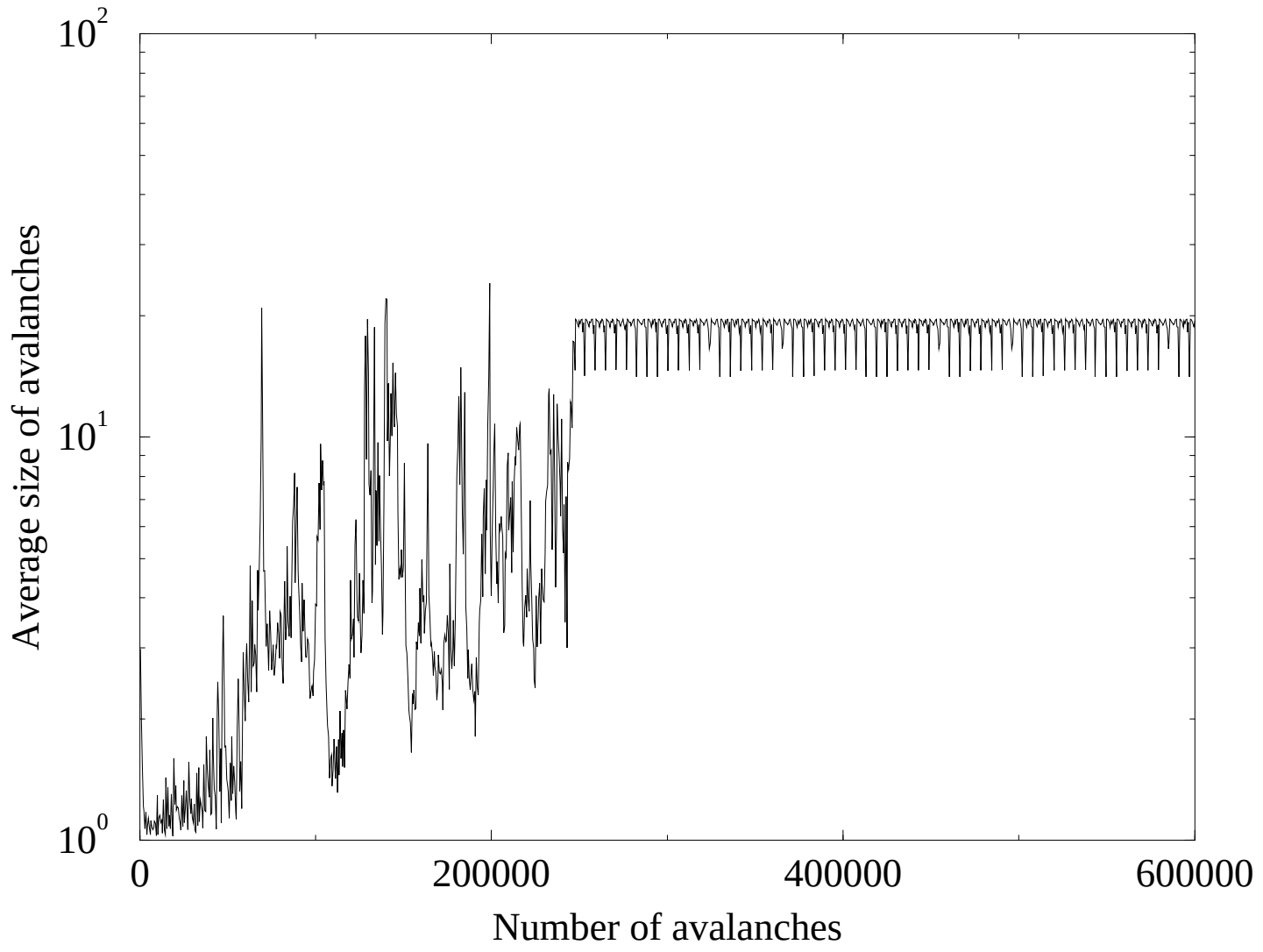

FIG. 4. Average size of the last 500 avalanches as a function of avalanche number for a $35 \times 35$ lattice with periodic boundaries and one doped site, $\alpha = 0.1825$ and $T^*_{\max} = 1.40$.

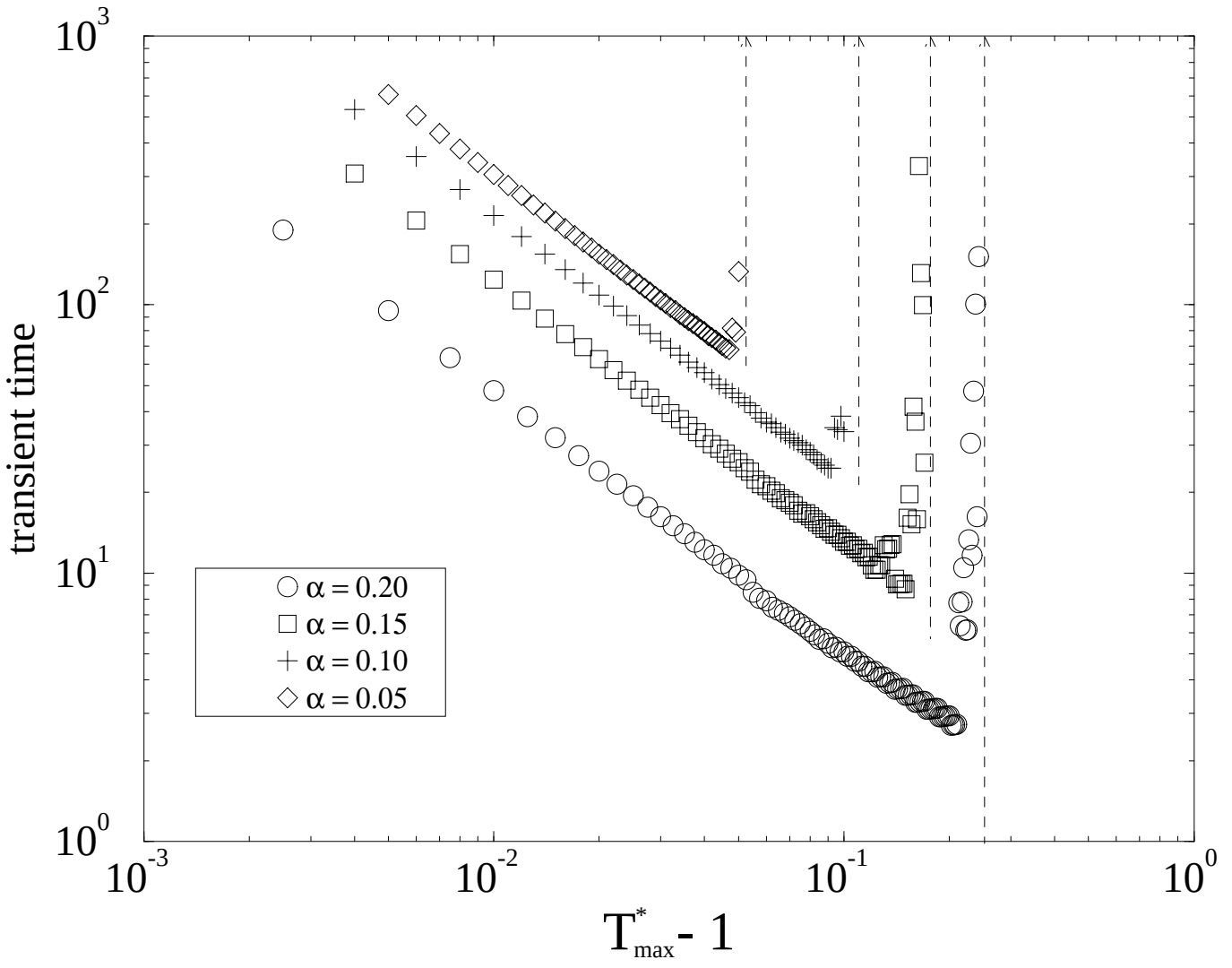

FIG. 5. Transient times for the strongly ordered state as a function of $T^*_{\rm max} - 1$ for different values of $\alpha$ and$L = 15$. The dashed lines indicate the values $T^*_{\rm max} = \frac{1}{1-\alpha}$ for the different $\alpha$.

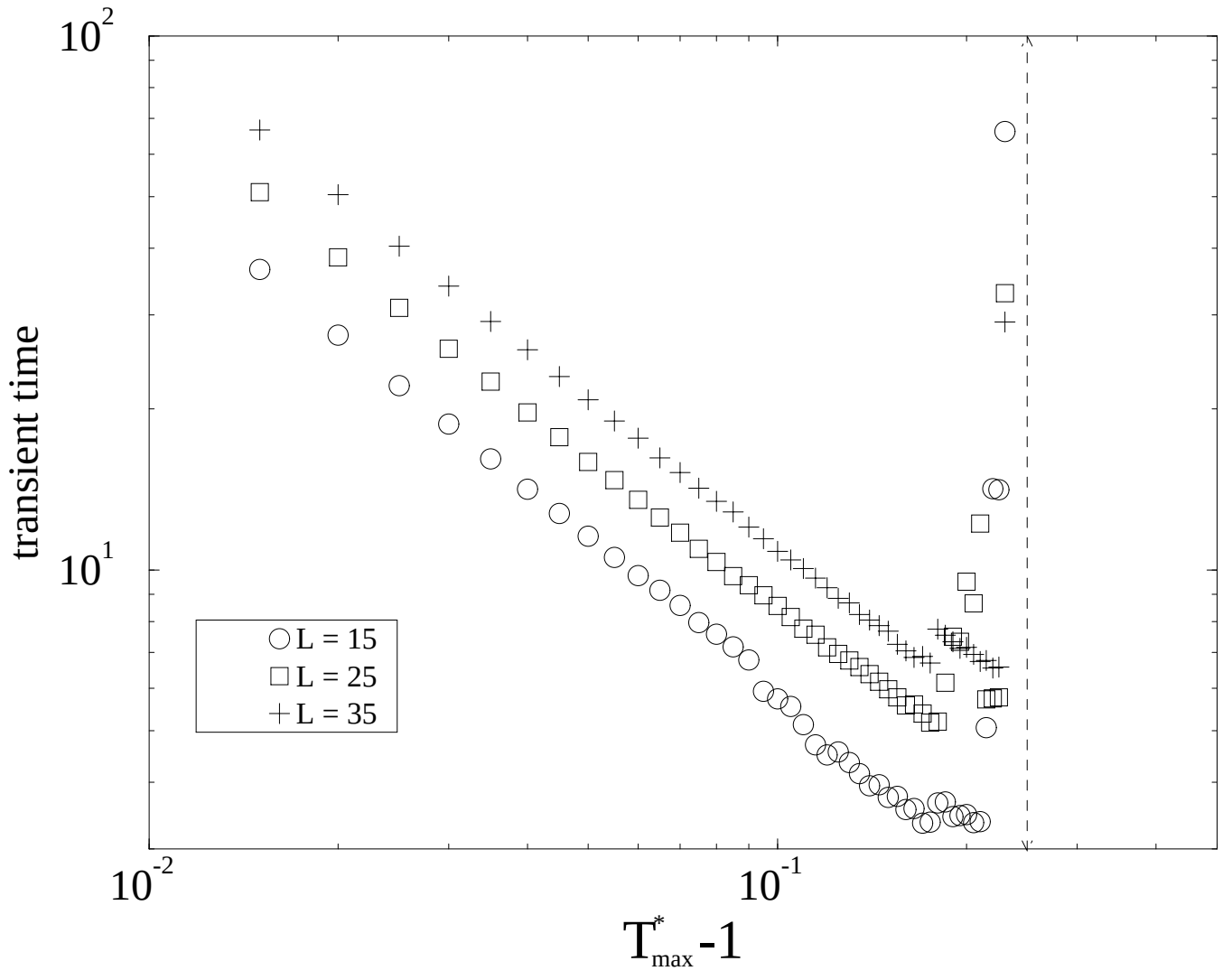

FIG. 6. Transient times for the strongly ordered state as a function of $T^*_{\rm max} - 1$ for different values of $L$ and $\alpha = 0.20$. The dashed line indicates the value $T^*_{\rm max} = \frac{1}{1-\alpha}$.

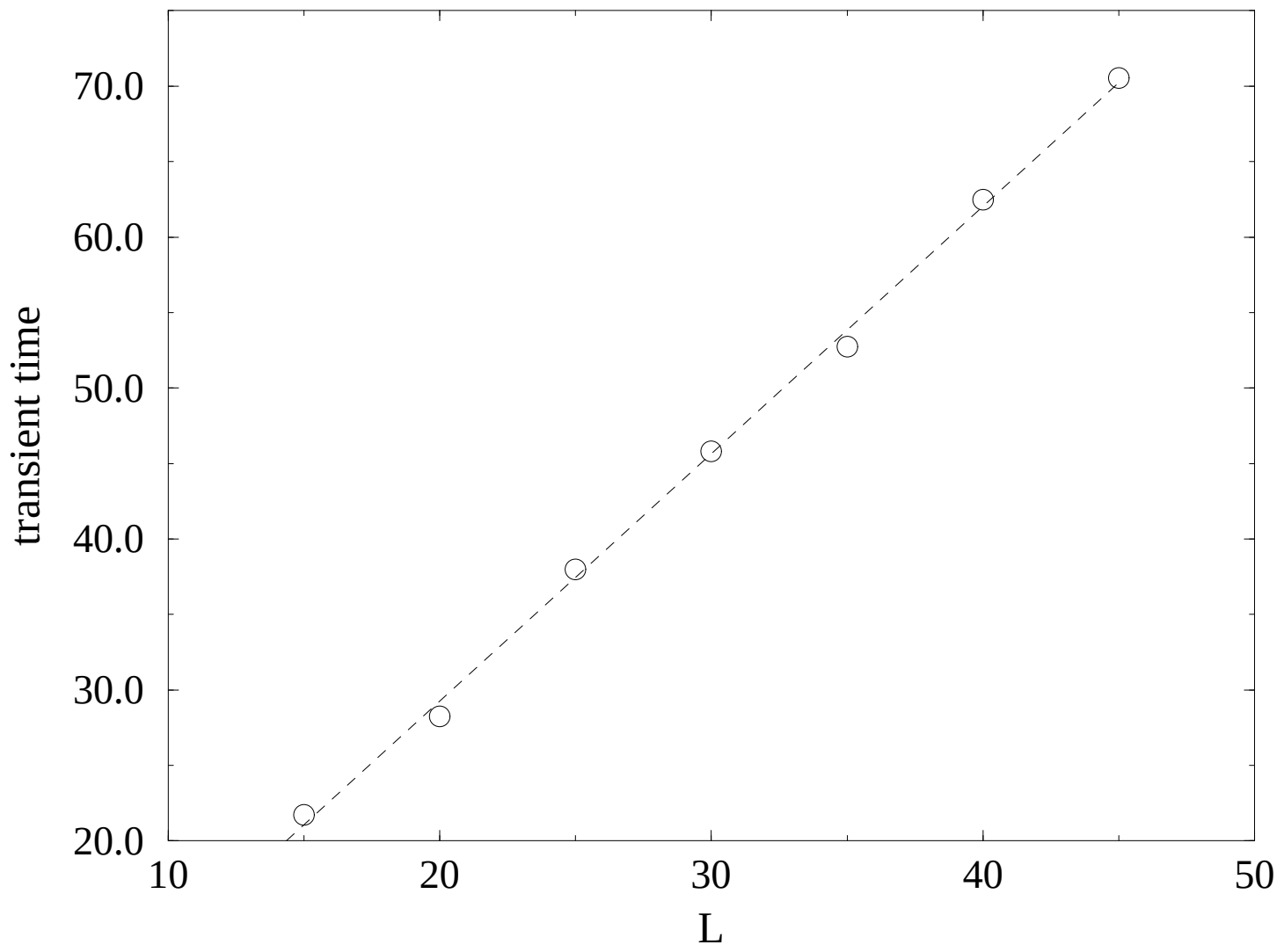

FIG. 7. Transient times for a lattice with $\alpha = 0.22$ and $T^*_{\rm max} = 1.01$ as a function of lattice size $L$. For each $L$, the mean value for 20 different initial conditions has been plotted. The data points are well fitted to a straight line with slope $\approx 1.6$.

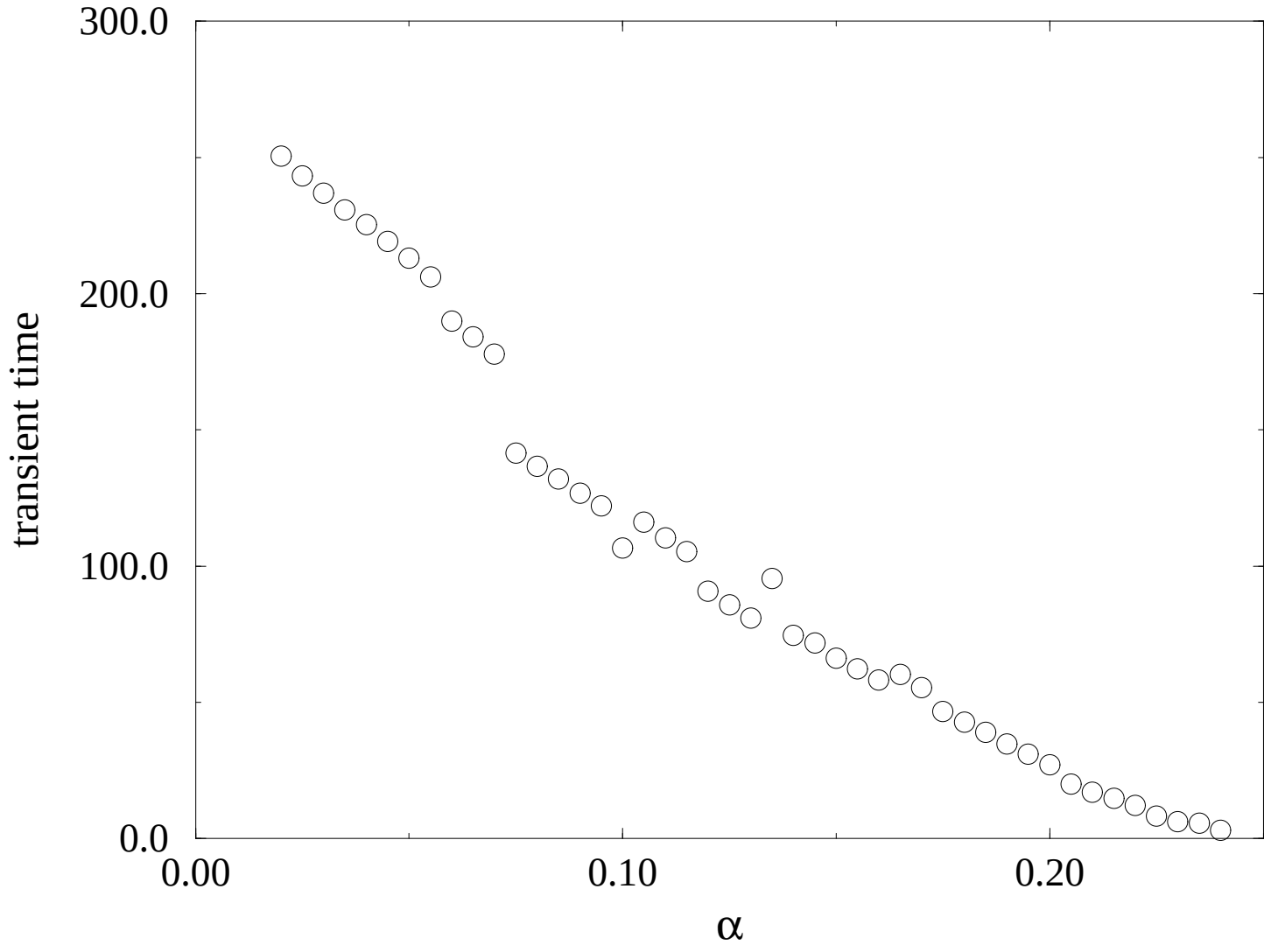

FIG. 8. Transient times as a function of $\alpha$ for $L = 15$ and $T^*_{\rm max} = 1.02$. The initial conditions are the same for all values of $\alpha$.